\documentstyle[editedvolume,epsf]{crckapb} 

\newcommand{\req}[1]{(\ref{#1})}              
\newcommand{\gcc}{{\rm~g\,cm}^{-3}}
\newcommand{\beq}{\begin{equation}}
\newcommand{\eeq}{\end{equation}}
\newcommand{\wc}{\omega_{\rm c}}
\newcommand{\am}{a_{\rm m}}
\newcommand{\mel}{m_{\rm e}}
\newcommand{\EF}{\epsilon_{\rm F}}
\newcommand{\GF}{\gamma_{\rm F}}
\newcommand{\vB}{\mbox{\boldmath{$B$}}}

\newcommand{\Compton}{\lambda\hspace{-.44em}\raisebox{.6ex}{\mbox{-$\!$-}}%
}
\renewcommand{\Compton}{\lambda\hspace{-.54em}{\mbox{$^-$}}%
}
\def\apj#1,#2{{\it Astrophys.\ J.}{\bf\ #1}, #2}
\def\apjs#1,#2{{\it Astrophys.\ J.\ Suppl.\ Ser.}{\bf\ #1}, #2}
\def\aa#1,#2{{\it Astron.\ \& Astrophys.}{\bf\ #1}, #2}
\def\aar#1,#2{{\it Astron.\ \& Astrophys.\ Rev.}{\bf\ #1}, #2}
\def\al#1,#2{{\it Astron.\ Lett.}{\bf\ #1}, #2}
\def\asr#1,#2{{\it Adv.\ Space Res.}{\bf\ #1}, #2}
\def\ass#1,#2{{\it Astrophys. Space Sci.}{\bf\ #1}, #2}
\def\mn#1,#2{{\it Mon.\ Not.\ R.\ Astron.\ Soc.}{\bf\ #1}, #2}
\def\nat#1,#2{{\it Nature}{\bf\ #1}, #2}
\def\pr#1,#2{{\it Phys. Rev.}{\bf\ #1}, #2}
\def\pre#1,#2{{\it Phys.\ Rev. E}{\bf\ #1}, #2}
\def\prl#1,#2{{\it Phys.\ Rev.\ Lett.}{\bf\ #1}, #2}

\def\la{\;\raise.5ex\hbox{$<$}\kern-.8em\lower 1mm\hbox{$\sim$}\;}
\def\ga{\;\raise.5ex\hbox{$>$}\kern-.8em\lower 1mm\hbox{$\sim$}\;}
\def\g{\gamma}
\def\h{\hbar}
\def\s{\sigma}
\def\sm{\simeq}
\def\ov{\over}

\def\<{\langle}
\def\>{\rangle}
\def\ri{\right}
\def\ea{{\it et al.} }

\def\ros{{\sl ROSAT} }
\begin{opening}
\title{NEUTRON STAR ENVELOPES AND THERMAL \protect\\
	 RADIATION FROM THE MAGNETIC SURFACE}

\author{Joseph Ventura}
\institute{Physics Department, University of Crete, and  
      IESL, FORTH\\
           71003 Heraklion, Crete, Greece}
\author{Alexander Y. Potekhin}
\institute{Ioffe Physico-Technical Institute, 194021 St.Petersburg, Russia} 

\end{opening}

\runningtitle{MAGNETIC NEUTRON STARS}

\begin{document}


\begin{abstract}
{\small
The thermal structure of neutron star envelopes is discussed with emphasis 
on analytic results. Recent progress on the effect of chemical
constitution 
and high magnetic fields on the opacities and the thermal structure is 
further reviewed in view of the application to pulsar cooling and
magnetars.
}
\end{abstract}

\section{Introduction}

Neutron stars (NS) are formed with very high internal temperatures
approaching 
$10^{11}$ K in the core of a supernova explosion (see, e.g., Shapiro \& 
Teukolski 1983 -- ST83 in the following). Copious neutrino emission brings
the 
temperature in the stellar core
down to $\simeq 10^9$ K within about one day and then, more 
gradually, to $\simeq 10^8$ K within $10^4$ years. It was realized early
on 
that such objects were likely to have effective surface temperatures of 
the order of $10^6$ K (Chiu \& Salpeter 1964). 
Comparison with theoretical cooling curves 
can further provide information on aspects of the internal 
structure of NS such as superfluidity and the possible appearance 
of pion or kaon condensates and strange matter in their interior
(e.g., Pethick 1992; Page 1997; Tsuruta 1998). It is then 
clear that the observation of this surface cooling X-ray 
emission is an objective of prime scientific importance. 

The observation of such faint point sources has turned out to be 
difficult however, having to await the modern 
era of imaging X-ray telescopes. 
\ros observations in the 90's finally yielded 
improved spectral information, 
opening a new chapter in our ability to probe the internal 
structure of superdense matter. Joachim Tr\"umper (this meeting) 
already gave us an overview 
of the recent observations in this field 
(see also Caraveo {\it et al.\/} 1996; Tr\"umper \& Becker 1997). 
Increasing interest is also presently 
being directed on the new and important subject of 
magnetars, extensively discussed during this meeting. 

Theoretical models of NS cooling are of necessity rather 
complicated, requiring detailed understanding of the stellar structure, 
equation of state (EOS), and thermal balance over an enormous range of 
densities and chemical diversity (e.g., Pethick 1992; 
Tsuruta 1998, and references therein). 
During an initial period 
of $~ 10^5 - 10^6$ y the star 
cools principally via neutrino emission from its interior. 
There are many neutrino emission mechanisms,
whose rates depend on the state of matter,
particularly on nucleon superfluidity 
(for a review see Yakovlev \ea 1999).
An older NS cools mainly via photon emission from its surface.
Finally, the cooling may also be influenced by 
heating processes such as friction due to differential rotation between 
the superfluid and normal parts of the star (e.g., 
Alpar {\it et al.\/} 1989), 
$\beta$-processes arising from chemical imbalance during the 
spindown (Reisenegger 1997), and pulsar
polar cap heating due to impinging charged particles accelerated 
in the magnetosphere (e.g., Halpern \& Ruderman 1993). 

Analysis of the NS thermal evolution 
is considerably simplified however by the 
fact that the stellar interior, 
from densities of $\simeq 10^{10}$ g cm$^{-3}$ inward, 
is nearly isothermal because of the very high thermal 
conductivity in these layers. 
It is therefore convenient 
to establish a relation between this interior temperature $T_{\rm i}$,
defined as the temperature at mass density $\rho=10^{10}$ g cm$^{-3}$, 
and the effective surface temperature $T_{\rm e}$ 
(Gudmundsson {\it et al.\/} 1983 -- GPE in the following). 
Thus, it becomes possible to examine 
separately the properties and structure of the outer envelope, which 
turns out to be 
crucial in determining the ratio $T_{\rm e}/T_{\rm i}$
and the nature of the emitted radiation 
(e.g., GPE; 
Hernquist \& Applegate 1984 -- HA84 in the following; Ventura 1989;
Potekhin {\it et al.\/} 1997 -- PCY in the following). 

In the next section, we consider basic features of the mechanical and
thermal
structure of the outer NS envelope without magnetic field. 
The strong magnetic field ($B\gg10^{10}$~G),
which was found in most of the pulsars,
shifts the atmosphere bottom and the region of partial 
ionization to higher densities. 
Furthermore, it strongly affects the radiative and thermal transport
through the envelope. These effects will be discussed
 in Sect.~3.

\section{Outer Envelope of a Neutron Star}
The {\em outer envelope\/} consists mainly of electrons and ions.
It extends down to 
a depth of a few hundred meters, 
where the density $4.3 \times 10^{11}$ g cm$^{-3}$
is reached.
At this density, 
neutrons begin to drip
from the nuclei (e.g., ST83).
Thus the {\em inner envelope\/}, which extends
deeper down to the core,
 consists of electrons, 
atomic nuclei, and free neutrons.

The outer envelope can be divided
into the atmosphere, the liquid ocean, and the solid crust.
The outermost layer constitutes a thin
(0.1 -- 100 cm) NS atmosphere
(optical depth $\tau \la 1$), where
the outgoing radiation is formed. 
The plasma density at the atmosphere bottom
is about 0.001 -- 0.1 g~cm$^{-3}$, depending on
temperature, surface gravity,
and chemical composition.
This plasma can be partially ionized and non-ideal.
Bound species can be distinct 
until the electron Fermi energy becomes comparable with the Thomas--Fermi energy
at $\rho \la 10 \, ZA$ g~cm$^{-3}$,
$A$ and $Z$ being the mass
and charge numbers.
At higher densities the ions are immersed in 
a jellium of degenerate electrons, 
which still strongly responds to the Coulomb 
fields of the ions as long as
$\rho\la 10 \, Z^2 A$ g cm$^{-3}$.

In the rest of the envelope (at $\rho\gg 10 \, Z^2 A$ g cm$^{-3}$)
the electrons form a strongly degenerate, almost ideal gas. 
This gas is non-relativistic at
$\rho_6 \ll 1$ and ultrarelativistic at
$\rho_6 \gg 1$, where $\rho_6\equiv\rho/10^6\gcc$. 
The ions form a Coulomb gas or liquid at $\Gamma\la175$, 
where $\Gamma\approx22.75\,(\rho_6/\mu_{\rm e})^{1/3}\,Z^{5/3}/T_6$
is the Coulomb coupling parameter, $T_6=T/10^6$~K,
$\mu_{\rm e}=A/Z$.
At $\Gamma\approx175$, 
the liquid freezes into a Coulomb crystal.
The pressure is almost entirely
determined by degenerate electrons and thus
independent of $T$, while the mass density
is mostly determined by the ions.

While cooling of the NS during the initial neutrino dominated era 
is not influenced by the outer layers, it is in fact the properties of 
these surface layers that characterize the flux and photon spectrum  
emitted at the NS surface, leading in turn to estimates of 
$T_{\rm i}$. 
 At the subsequent photon cooling stage, 
the heat insulation by the envelope controls the cooling rate.
The radiation is reprocessed 
in the atmosphere, which yields
an emitted spectrum, in general
different from that of a black body
(Romani 1987).
The properties of these layers are thus 
crucial to interpreting observations and, understandably, 
a lot of theoretical work has been devoted to analyzing the 
thermal structure of non-magnetic and magnetic NS envelopes
(e.g., GPE; 
PCY; Heyl \& Hernquist 1998) and
the radiation properties of their atmospheres  
(e.g., Pavlov {\it et al.\/} 1995;
Page \& Sarmiento 1996; Rajagopal {\it et al.\/} 1997; 
Potekhin \ea 1998, and references therein). 

\subsection{Mechanical structure}
To review the cardinal properties of the NS surface layers, 
let us recall the enormous gravitational potential 
$GM/R \approx 0.148 \,(M/M_\odot)\,R_6^{-1}\,c^2$
(where $R_6\equiv R/10^6{\rm~cm}\sim1$ and $M/M_\odot\sm1.4$ for
most typical NS), which renders effects
of General Relativity appreciable.
Indeed, the Schwarzschild radius $r_g=2GM/c^2\approx2.95\,(M/M_\odot)$~km
is not much smaller than the stellar radius $R$
(throughout this review, $M$ is the gravitational mass of the star,
which is 10--15\% smaller than its baryon mass).
The hydrostatic equilibrium is then governed by
the Oppenheimer--Volkoff equation (e.g., Thorne 1977).
Introducing the local proper depth $z=(R-r)\,(1-r_g/R)^{-1/2}$
(where $r$ is the radius), in the surface layers ($z\ll R$)
one can rewrite this equation in the Newtonian form 
\begin{equation}
{\rm d}P/{\rm d}z = g\rho, 
\label{hydrostat}
\end{equation}
where
$$
g={GM\ov R^2\,(1-r_g/R)^{1/2}}
 \approx 1.327\times10^{14}\,(1-r_g/R)^{-1/2}\,{M\ov M_\odot}\,R_6^{-2}
{\rm~cm~s}^{-2}
$$
 is the local gravitational acceleration at the surface.

 Since the surface gravity is huge,
 the atmosphere's scale height is rather small.
In the non-degenerate layers, we have 
\begin{equation}
   P=(\rho/\mu m_{\rm u})\,kT,
\label{P-nondeg}
\end{equation}
where  $m_{\rm u}=1.6605\times10^{-24}$~g 
is the atomic mass unit and $\mu =A/(Z+1)$.
Thus, following a thin non-degenerate atmosphere of a scale height
$ P/(g \rho) \approx 0.626\,
({T_6} / \mu )\,R_6^2\,({M / M_\odot})^{-1}$ cm,
electron 
degeneracy sets in at densities $\rho \ga 6\,\mu_{\rm e} T_{6}^{3/2}$ g cm$^{-3}$.

The electron kinetic energy at the Fermi surface is 
$kT_{\rm F}\equiv\EF-\mel c^2$,
where $k$ is the Boltzmann constant,
\begin{equation}
 T_{\rm F}= \left(\g_{\rm F}-1\ri)\mel c^2/k= 
5.93\times10^9\;{\chi^2}/(1+\g_{\rm F}){\rm~K}
 \label{T_F}
\end{equation}
 is the Fermi temperature, and
\begin{equation}
  \g_{\rm F}=\sqrt{1+\chi^2},
\qquad
 \chi= {p_{\rm F}\,\ov {\mel c}}= 
 {{\h(3\pi^2n_{\rm e})^{1/3}}\ov{\mel c}}\approx
 {1.009 \left({\rho_6\ov \mu_{\rm e}}\ri)^{1/3}},
\label{x_r}
\end{equation}
are the electron Lorentz factor and {\em relativity parameter}, respectively.

Elementary fitting formulae to the pressure of fully ionized
ion-electron plasmas as function of density 
at arbitrary electron degeneracy and temperature
have been presented by Chabrier \& Potekhin (1998).
These formulae can also be extended to
partially ionized atmospheric layers in the mean-ion approximation,
provided the effective ion charge $Z$ is known
(cf.\ PCY).
At sufficiently high density, however, where $T_{\rm F}\gg T$, 
the main contribution is that of strongly degenerate electrons with 
the pressure depending 
only on the density through the parameter $\chi$
(e.g., ST83):
\begin{equation}
 P_{\rm e}=
 {P_0\over8\pi^2}\,\left[\,\chi\left(\frac23\,\chi^2-1\right)\GF
             + \ln(\chi+\GF)\right]
\label{P-d}
\end{equation}      
where $P_0 \equiv \mel c^2/\Compton^3
 = 1.4218\times 10^{25}{\rm~dynes~cm}^{-2}$
is the relativistic unit of pressure,
$\Compton=\hbar/(\mel c)$ being the Compton wavelength.
This may further be approximated as 
$
   P_{\rm e} \approx 
      P_0 \chi^{3\gamma_{\rm ad}}/(9\pi^2\gamma_{\rm ad}),
$
where $\gamma_{\rm ad}$ is the adiabatic index,
equal to 5/3 at $\chi\ll1$ and 4/3 at $\chi\gg1$. 

Note that for strongly degenerate electrons
\beq
{\rm d}P=n_{\rm e}{\rm\,d}\EF = 4.93\times10^{17}{\rm~erg~g}^{-1}\,
(\rho/\mu_{\rm e})\,\chi{\rm\,d}\chi/\gamma_{\rm F}.
\label{dP}
\eeq
{}From Eqs.~\req{hydrostat} and \req{dP},
one obtains
\begin{equation}
1.027\,{\rho_6\ov \mu_{\rm e}} = 
\chi^3= \left[{z \over z_0} \left( 2 + {z \over z_0} \right)\right]^{3/2},
   \quad
   z_0= {\mel c^2 \over m_{\rm u} g \mu_{\rm e}}\approx 
       {4930 {\rm~cm}\over \mu_{\rm e}\, g_{14}},
\label{str-z}
\end{equation}
where $g_{14}=g/(10^{14}\;{\rm cm \, s^{-2}})$,
and $z_0$ is a depth scale at which degenerate electrons
become relativistic.

Let us note also that  the mass $\Delta M$  contained in
a layer from the surface to a given depth $z$ 
is solely determined by the
pressure $P(z)$ at the bottom of the given layer:
\begin{equation}
     \Delta M(z)= 4 \pi R^2 \, P(z) \,g^{-1} \, (1- r_g/R)^{1/2}
\approx 1.192\times10^{-9}\,g_{14}^{-2}\,MP(z)/P_0.
\label{str-dM}
\end{equation}

\subsection{Thermal Structure}
{}From Eqs.~\req{str-z} and \req{str-dM}, we see that
 the outer envelope is a very thin layer -- 
typically within the outer 100 m, containing $\sim 10^{-7}$M$_\odot$, --
which renders the thermal diffusion problem essentially 
plane-parallel and one-dimensional. 
Assuming a constant heat flux throughout the outer envelope, 
the temperature profile 
can be obtained by solving the heat diffusion equation:
\begin{equation}
 F= \kappa {{\rm\,d}T\ov{\rm\,d}z}=
      {16\sigma\over3}{T^3{\rm\,d}T\over{\rm\,d}\tau},
\quad
    \kappa\equiv{16\sigma T^3\over3K\rho},
\label{Flx}
\end{equation}      
where $F$ is the heat flux, 
$\kappa$ is thermal conductivity, 
 $\sigma$ is
the Stefan--Boltz\-mann constant, and
$K$ is the usual Rosseland mean over the energy spectrum 
of the specific opacity. 
This leads immediately to a temperature profile  
$
 {T/{T_{\rm e}}}\approx (\frac34\tau+\frac12)^{1/4},
$      
where the {\em local\/}
effective surface temperature $T_{\rm e}$ is defined through 
$F=\s T_{\rm e}^4$
and the integration constant corresponds to the Eddington 
approximation ($\tau=\frac23$ at the {\em radiative surface\/},
where $T=T_{\rm e}$).
A more accurate boundary condition 
requires solution of the radiative transfer equation in 
the atmosphere.
A distant observer would infer from the spectrum and flux
the redshifted surface temperature $T_\infty=T_{\rm e}\,(1-r_g/R)^{1/2}$ 
and apparent radius $R_\infty=R\,(1-r_g/R)^{-1/2}$ (e.g., Thorne 1977).

A knowledge of the mean opacity 
$K(\rho,T)$ is then needed to relate the temperature to the other 
plasma parameters, which can then also be expressed as functions of 
the optical depth $\tau$. It is also needed  
in order to compute the overall depth 
and the temperature ratio $T_{\rm i}/T_{\rm e}$.

\subsubsection{Opacity} 
Heat is transported through the envelope mainly by radiation and by 
conduction electrons. In general,
 the two mechanisms work in parallel, hence
\begin{equation}
\overline\kappa=\kappa_{\rm r}+\kappa_{\rm c},
\qquad
 K^{-1} = {K}_{\rm r}^{-1} + {K}_{\rm c}^{-1}, 
 \label{op}
\end{equation}      
where $\kappa_{\rm r}$, $\kappa_{\rm c}$ 
and ${K}_{\rm r}$, ${K}_{\rm c}$ denote the radiation and conduction 
components of the conductivity and opacity, respectively. 
Typically, the radiative conduction dominates the thermal transport 
($\kappa_{\rm r}>\kappa_{\rm c}$) in 
the outermost non-degenerate layers of a NS,
whereas electron conduction dominates ($\kappa_{\rm c}>\kappa_{\rm r}$)
in deeper, mostly degenerate layers.
In the absence of intense magnetic fields, modern cooling calculations 
(e.g., PCY) make use of the Livermore library of radiative opacities 
{\sc opal} (Iglesias \& Rogers 1996),
which also provides 
an EOS for the relevant thermodynamic parameters 
at $\rho\la 10\,T_6^3\gcc$.
For the electron conduction regime, modern opacities
have been worked out by Potekhin \ea (1999a).

\paragraph{Radiative opacities.}
In order to derive an analytic model of the NS envelope,
HA84 and Ventura (1989) have written the atmospheric opacity in the form
\beq
K(\rho,T) = K_0 \rho^\alpha\,T^\beta.
\label{K_rad}
\eeq
In particular, this relation describes the opacity given by the
Kramers formula,
which corresponds to $\alpha=1$ and $\beta=-3.5$.
In a fully
ionized, non-relativistic and
non-degenerate
plasma, the opacity provided by the
free-free transitions is (e.g., Cox \& Giuli 1968)
\beq
 K_{\rm r} \approx75\,\bar{g}_{\rm eff}\,(Z/\mu_{\rm e}^2)\,\rho\,T_6^{-3.5}
{\rm~{cm^2/ g}},
\label{K_0}
\eeq
where $\rho$ is in g~cm$^{-3}$ and 
$\bar{g}_{\rm eff} \sim 1$ is an effective dimensionless Gaunt factor,
a slow function of the plasma parameters. 
For a colder
plasma, where bound-free transitions dominate over
free-free ones, the Kramers formula remains approximately valid, but the 
opacity $K$ is about two orders of magnitude
higher.
An order-of-magnitude (within $\approx0.5$ in $\log K$) approximation to
the realistic {\sc opal} opacities for hydrogen 
in the range of parameters $T_6\sim10^{-1}-10^{0.5}$
and $\rho\sim(10^{-2}-10^1)\,T_6^3$
is given by Eq.~\req{K_0} if we put formally 
$\bar{g}_{\rm eff}\approx\rho^{-0.2}$.
An analogous order-of-magnitude approximation to
the {\sc opal} opacities for iron at $T_6\sim10^0-10^{1.5}$
and $\rho\sim(10^{-4}-10^{-1})\,T_6^3$
is obtained with $\bar{g}_{\rm eff}\approx70\,\rho^{-0.2}$.
These approximations also belong to the class
of functions (\ref{K_rad}), but with $\alpha=0.8$.

\paragraph{Conductive opacities.}
Thermal conduction of degenerate matter in deep\-er layers
is dominated by electrons which scatter off ions.
This conductivity can be written as (Yakovlev \& Urpin 1980) 
\beq
     \kappa_{\rm c} = { \pi k^2 T \mel c^3 \chi^3
             \over
             12 Z e^4 \Lambda \GF^2}
\approx 2.3\times10^{15}\,{T_6\ov\Lambda Z}\,{\chi^3\ov1+\chi^2}
{\rm~{erg\ov cm~s~K}},
\label{K_ei}
\eeq
where $\Lambda$ is the Coulomb logarithm.
Accurate analytic fitting formulas to $\Lambda$ as function of $\rho$
and $T$ have been obtained recently by Potekhin {\it et al.\/} (1999a).
In the solid crust, 
this function is reduced to small values by quantum 
and correlation effects in Coulomb
crystals. However, we shall see shortly that
in not too cold NS, the thermal profile is mainly
formed in the liquid layers
of the envelope. Therefore,
for our purpose, it will be sufficient to note that 
in the NS ocean $\Lambda$ is a slow
function of the plasma parameters and can be approximated by a constant of
the order of unity.

In the case of non-degenerate electrons, the conductivity
can be found, e.g., by the method of Braginski\u\i\ (1957),
which yields 
\beq
     \kappa_{\rm c}^{\rm nd} 
\approx 5\times10^{10}\,(F_Z/\Lambda)\, Z^{-1} \, T_6^{5/2}
{\rm~erg/(cm~s~K)},
\label{kappa-ND}
\eeq
where $F_Z$ is a slow function of $Z$: for example,
$F_{26}=1.34$ and $F_1=0.36$,
whereas the Coulomb logarithm $\Lambda$ is $\sim1$
near the onset of degeneracy
and logarithmically increases with decreasing density. 
Effectively, Eq.~\req{kappa-ND} may be viewed as an analog  
to Eq.~\req{K_ei} where the dimensionless 
Fermi momentum $\chi$
has been replaced by an appropriate thermal average ($\propto\sqrt{T}$). 

\subsubsection{Temperature Profile}
The simple functional form of the opacity  allows now an analytic treatment of
the thermal structure (Urpin \& Yakovlev 1980; HA84; Ventura 1989).

\paragraph{Non-degenerate regime.}
Using Eqs.~\req{hydrostat}, \req{K_rad}, and \req{P-nondeg}
one may rewrite Eq.~\req{Flx} as 
\begin{equation}
 {\rm d}P={\kappa\,g\,\rho\ov F}{\rm\,d}T
  ={16\over3}\,{g\over K}\,{T^3{\rm\,d}T\over T_{\rm e}^4}
=
{16\ov 3}\,{g\ov K_0}
\left(k\ov {\mu  m_{\rm u}}\ri)^\alpha {T^{3+\alpha-\beta}\ov {P^\alpha }}\,
{{\rm d}T\ov{T_{\rm e}^4}}
  ,
  \label{dpdt}
\end{equation}      
which 
 is readily
integrated from the surface inward to give the temperature profile.
In the region far from the surface, 
where $T\gg T_{\rm e}$,
the integration constants
may be dropped. 
Using again Eqs.~\req{P-nondeg}, one 
can present the result as a simple power law:
\begin{equation}
 {T^{3-\beta} \ov \rho^{\alpha+1}}= {3\ov16}\,
    {{4+\alpha-\beta}\ov {\alpha+1}}\,{k\over \mu  m_{\rm u}}
      \,{K_0 T_{\rm e}^4\ov g}.
  \label{prlw}
\end{equation} 
This result depends on our implicit assumption that 
the effective ion charge $Z$ remains constant, which in general 
is not strictly valid.

Interestingly enough, as noted by HA84, 
Eq.~\req{prlw} establishes that the conductivity $\kappa$ 
is constant throughout the radiative non-degenerate layer:
\begin{equation}
 \kappa ={16\ov 3} {{\s T^3}\ov {K\rho}}= 
  {16\ov 3}{\s\ov K_0} {T ^{3-\beta} \ov \rho^{\alpha+1}}=  
  {{4+\alpha-\beta}\ov {1+\alpha}}{{k\s T_{\rm e}^4}\ov {\mu  m_{\rm u} g}}.
\label{k-const}
\end{equation} 
The constant value depends on the emitted flux, 
but is independent of $K_0$.
It follows immediately from Eq.~\req{Flx}
that temperature grows linearly with geometrical depth.
Furthermore, from Eq.~\req{prlw} we obtain the 
$T,\rho$ profile.
In particular, substituting $\alpha=1$ and $\beta=-3.5$,
we obtain 
\beq
  K\rho={1.51\ov{\rm cm}}\,{\mu g_{14}\ov T_{\rm e6}}
     \left({T\over T_{\rm e}}\right)^3,
\quad
\kappa = 2\times10^{14} \, {T_{\rm e6}^4\over \mu  g_{14}}
{\rm\,{erg\over cm\,s\,K}}.
\eeq
Substitution of $K$ from Eq.~\req{K_0} yields
\beq
T_6 \approx 0.284 \, g_{14}\,\mu \, z
\approx (50\,\bar{g}_{\rm eff}\,q)^{2/13}\,(\rho/\mu_{\rm e})^{4/13},
\quad 
q\equiv T_{\rm e6}^4\,Z/(\mu g_{14}).
\label{nd-solution}
\eeq
In these relations, $z$ and $\rho$ are measured in CGS units,
and $T_{\rm e6}=T_{\rm e}/10^6$~K.

\paragraph{Radiative surface.} It is further interesting to 
note that  the opacity $K$ is also slowly varying in this
region. Combining Eqs.~\req{K_rad}, and 	
\req{nd-solution}, we obtain $K\propto \rho/T^{3.5}\propto \rho^{-1/13}$. 
Invoking Eq.~\req{P-nondeg}, we get $K\propto P^{-1/17}$. 
This justifies the assumption 
$K_{\rm r}\approx$ const,
which one often employs when determining the radiative
surface from equation  (e.g., GPE; PCY)
\beq
(K_{\rm r} P)_{\rm surface} = (2/3)\, g.
\label{KP}
\eeq
Using Eqs.~\req{P-nondeg}
and \req{K_0}, we obtain
\beq
    \rho_{\rm s} \approx 0.1\,\mu_{\rm e}\left({\mu \ov Z\,
\bar{g}_{\rm eff}}\,g_{14}\right)^{1/2} T_{\rm e6}^{5/4} \gcc.
\label{surface}
\eeq
Substituting $\bar{g}_{\rm eff}\sm1$ and $\bar{g}_{\rm eff}\sm200$
for hydrogen and iron, respectively,
we obtain $\rho_{\rm s}\sim0.07\,\sqrt{g_{14}}\,T_{\rm e6}^{5/4}\gcc$
and  $\rho_{\rm s}\sim0.004\,\sqrt{g_{14}}\,T_{\rm e6}^{5/4}\gcc$
for these two elements, in reasonable agreement
with numerical results of PCY.

\paragraph{Onset of degeneracy.} 
The solution given by Eq.~\req{nd-solution}
can be extended down to a depth
where the electrons become degenerate. 
 Let us estimate this
depth from the condition $k T_{\rm d}= p_{\rm F}^2/2\mel$. 
We obtain
\beq
   \chi_{\rm d} \simeq 0.053 \,(\bar{g}_{\rm eff}\,q)^{1/7},
\quad
   T_{\rm d} \simeq 8.5 \times 10^6 \,(\bar{g}_{\rm eff}\,q)^{2/7}{\rm~K}.
\label{degenerate}
\eeq
 Even for very high $T_{\rm e} \sim 10^7$~K, we have
$\chi_{\rm d} \la 1$, i.e.\ the electrons are non-relativistic 
at the degeneracy boundary. With decreasing $T_{\rm e}$,
the quantities $\chi_{\rm d}$ and $T_{\rm d}$
decrease, i.e.\ the boundary shifts toward the stellar
surface.

\paragraph{Sensitivity strip.}
Numerical cooling calculations (GPE) revealed early on that
accurate knowledge of the opacity law is especially important within a certain 
``sensitivity strip'' in the $(\rho,T)$ plane. 
The temperature ratio $T_{\rm i}/T_{\rm e}$ 
changes appreciably if $K$ is modified by, say, a factor 
2 within this narrow strip, while comparable changes of  $K$ 
outside the strip would leave this ratio unaffected within 
a high degree of accuracy. 

The importance of the layer in the outer envelope where the opacity \req{op}
turns from radiative to conduction dominated is now easy to demonstrate.
In the non-degenerate radiative part the integral over $\rho$ is dominated 
by the higher densities near the base of the layer, while in the degenerate, 
conductive layer it is dominated by the top, least dense part 
of the layer. The region where 
$K$ turns from radiative to conductive is thus seen to
contribute most of the resistance to heat flow. 
The line in the $(\rho,T)$ plane where ${K}_{\rm r}={K}_{\rm c}$
 is easily determined from Eqs.~\req{K_0} and \req{K_ei}:
\beq
   \rho\approx 12\,\mu_{\rm e}\bar{g}_{\rm eff}^{-1/3}\, T_6^{11/6}\gcc.
\label{turn1}
\eeq
On the right-hand side, we have approximated the factor 
$(\Lambda\GF^2)^{1/3}$ by unity.
Using again the solution \req{prlw}, we find explicitly
the temperature $T_{\rm t}$ and relativity factor $\chi_{\rm t}$
at the turning point from radiative to electron conduction:
\beq
   T_{\rm t}\approx 2.3\times10^7\,\bar{g}_{\rm eff}^{2/17}\,
     q^{6/17}{\rm~K},
\quad
   \chi_{\rm t}\approx 0.157\,\bar{g}_{\rm eff}^{-2/51}
     q^{11/51}.
\label{turn}
\eeq
Some caution is necessary here, however, because the approximation 
$K\sm {K}_{\rm r}$ is not justified as we approach the turning point: 
actually, at this point $K=K_{\rm r}/2$, as seen from Eq.~\req{op}. 
In addition, the extrapolation of the solution \req{nd-solution}
to the turning point is, strictly speaking, not justified,
since the electron gas becomes degenerate:
$\chi_{\rm t}>\chi_{\rm d}$ for most typical NS parameters. 
Nevertheless,
since $\chi_{\rm t}$ and $\chi_{\rm d}$ are not very
much different, the section of the thermal profile where
our assumptions are violated is relatively small, so
Eq.~\req{turn}
provides a reasonable approximation. 
This is confirmed by a direct comparison with numerical results (PCY),
which reveals an error 
within only a few tens percent, provided $T\ga10^{5.5}$~K.

\paragraph{Solution beyond the turning point.}
An analytic solution to the thermal profile
in degenerate layers of a NS envelope has been first obtained by
Urpin \& Yakovlev (1980), 
based on the conductivity in the form \req{K_ei}.
The hydrostatic equilibrium of the degenerate surface layers
is determined by Eqs.~\req{hydrostat} and \req{dP},
which yield $g m_{\rm u}\,\mu_{\rm e} =
 \mel c^2 (\chi/\GF) \, {\rm d} \chi / {\rm d} z $.
Using the heat diffusion equation \req{Flx} and Eq.~\req{K_ei},
 we obtain
\beq
    T \, {{\rm d} T \over {\rm d} \chi}={12\over\pi}\,
    { F Z  e^4 \Lambda \over m_{\rm u} k^2 c g \mu_{\rm e}}\,
    {\GF \over \chi^2 }
=(1.56\times10^7{\rm~K})^2\,{Z\Lambda T_{\rm e6}^4\over \mu_{\rm e} g_{14}}\,
    {\GF \over 2\chi^2 }.
\label{dT/dx}
\eeq
Treating $\Lambda$, $A$, and $Z$ as constants, we can integrate
this equation from $\chi_{\rm t}$ inwards and obtain
\beq
   T^2(z)  = T_{\rm t}^2+
 (1.56\times10^7{\rm~K})^2\,{Z\Lambda T_{\rm e6}^4\over \mu_{\rm e} g_{14}}
      \left[f(\chi)-f(\chi_{\rm t})\right],
\label{T1(z)}
\eeq 
where $f(\chi) \equiv \ln(\chi+\GF)-\GF/\chi$,
and the dependence of $\chi$ on $z$ is given by Eq.~\req{str-z}.

Let us use the above solution to evaluate $T_{\rm i}$.
 Since $\chi_{\rm t}\la 1$, but $\chi\gg1$, it is easy to see that
 $[f(\chi)-f(\chi_{\rm t})]\approx\chi_{\rm t}^{-1}$ 
in Eq.~\req{T1(z)}.
Using Eq.~\req{turn}, we obtain
\beq
T_{\rm i} = \left(T_{\rm t}^2+T_{\Delta}^2\right)^{1/2},
\quad
T_{\Delta}\approx 4 \times 10^7 \left( { Z\,T_{\rm e6}^4 \over \mu_{\rm e} g_{14}}
    \right)^{20/51}{\rm~K},
\label{T1}
\eeq
where we have neglected some factors which are close to unity.

\begin{figure}[t]
\epsfysize=80mm
\epsffile[30 245 565 590]{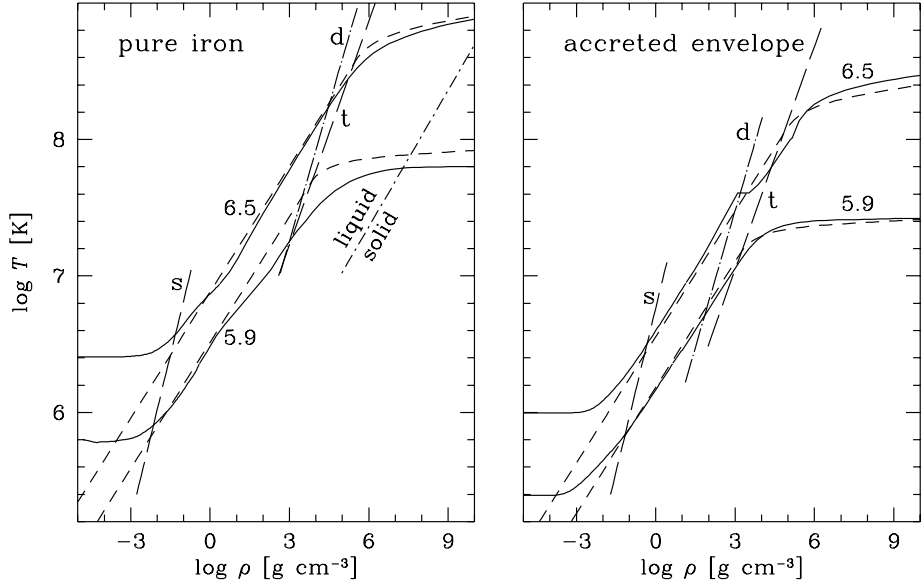}
\caption{Thermal profiles inside non-accreted (left panel)
and accreted (right panel) envelopes of NS at two effective temperatures,
$\log T_{\rm e}{\rm~[K]}=5.9$ and 6.5 (marked near the curves).
Solid curves -- numerical solution (Shibanov \ea 1998),
dashed curves -- analytic approximations \protect\req{nd-solution} and
\protect\req{T1}.
Straight lines marked ``s", ``d," and ``t" give the values 
$\rho$ and $T$ at which the various temperature profiles
cross the radiative surface [Eq.~\protect\req{surface}], 
the onset of electron degeneracy [Eq.~\protect\req{degenerate}],
and the turning point  [Eq.~\protect\req{turn}], respectively.
The Fe melting line is also shown.}
\label{fig-profiles}
\end{figure}

Figure \ref{fig-profiles} illustrates the accuracy of the above analytic
solution as well as its limitations.
Here, solid lines show temperature profiles of a NS 
with mass $M=1.4\,M_\odot$ and radius $R=10$ km
obtained by solving the radiative transfer equation with spectral {\sc opal}
opacities in the atmosphere and integrating the thermal structure
equation \req{Flx} with accurate radiative and conductive
Rosseland opacities inwards in the deeper layers (Shibanov \ea 1998).
Dashed curves depict the above analytic approximations.
The left panel corresponds to an envelope composed of iron,
and the right panel shows the thermal structure of an accreted envelope
with its outermost layers composed of hydrogen,
which is further burnt into heavier elements (He, C, Fe)
in deeper and hotter layers. 
This thermo- and pycnonuclear burning is responsible
for the complex shape of the upper profile.
Different straight lines
show the points at which thermal profiles corresponding to various 
heat fluxes would 
cross the radiative surface, the region of the onset of degeneracy,
turning points $K_{\rm c}=K_{\rm r}$, and (on the left panel)
the bottom of the ocean. The latter line is not present on the right panel,
because freezing of hydrogen and helium is suppressed by
large zero-point ionic vibrations (cf.\ Chabrier 1993).

One can see that not only the above analytic approximations
correctly describe the qualitative structure of the envelope,
but they also provide a reasonable quantitative estimate
of the temperature at a given density.
The accuracy deteriorates for lower $T_{\rm e}$ (especially
when $Z$ is high), because in this case radiative opacities
are affected by bound-bound transitions
and strong plasma coupling effects and thus no longer obey the simple
power law \req{K_0}.

\paragraph{Discussion.}
We have seen that the internal temperature $T_{\rm i}$
is determined by two temperatures, $T_{\rm t}$ and $T_{\Delta}$,
related to the
non-degenerate and degenerate layers, respectively,
and that the temperature growth occurs in
the very surface layers of the star. This justifies the
separation of the NS into a blanketing envelope and
internal isothermal layers used in numerical simulations (GPE; PCY).
We have also shown, in agreement with GPE,
 that the bulk of the envelope's thermal insulation arises 
in the relatively low density region around the turning point 
defined by Eq.~\req{turn}. 

The dependence on $Z$ and $\mu_{\rm e}$ that enters Eqs.~\req{T1}
makes the $T_{\rm i}/T_{\rm e}$ 
ratio sensitively dependent of the chemical composition 
of the envelope. If we replace the iron envelope by
an accreted envelope composed of hydrogen or helium,
we get this ratio reduced by about a half order of magnitude,
which corresponds to two orders of magnitude higher
photon luminosity at a given internal temperature.
Thus an envelope composed of light elements is much more transparent
to heat, and this strongly affects cooling, as
noted by PCY. 
Our analytic estimates present a tool for the 
fast estimation of the magnitude of such effects.

In the next section we will see how these principal properties 
change as a result of a strong magnetic field permeating 
the NS envelope.  

\section{Effects of Magnetic Fields}

It is well known that the strong magnetic field can profoundly 
alter the physical properties of the NS outer layers --
for reviews see Canuto \& Ventura (1977), M\'esz\'aros (1992), and
Yakovlev \& Kaminker (1994).
We will be interested here in modifications introduced by the 
intense field $B$
in the EOS, the Fermi temperature $T_{\rm F}$, and the opacity
$K$, all of which affect the heat transfer problem. 
Our primary focus will be on the properties of an electron gas.
Free ions give only a minor contribution to the opacities,
whereas their contribution to the EOS remains unaffected by the field
in the non-degenerate regime and is negligible when electrons are
strongly degenerate.

Apart from the bulk properties of the electron plasma, 
magnetic fields also modify profoundly the properties of 
atoms which become very elongated and compact, having sharply increased 
binding energies (e.g., Canuto \& Ventura 1977). 
This should strongly affect the emitted spectra 
from NS surfaces. 
Many works have been devoted in the past to the calculation
of quantum-mechanical properties of atoms at rest in strong magnetic
fields (e.g., Miller \& Neuhauser 1991). Some of them have been used
to construct magnetic NS atmosphere models (Rajagopal \ea 1997).
However, thermal motion of the atoms at realistic NS temperatures
breaks down the axial symmetry and may completely alter
atomic properties.
It is therefore necessary to have 
quantum-mechanical calculations for {\it moving\/}
atoms, and their results should be included in models of EOS and opacities.
Such models are available to date only for hydrogen (e.g., 
Potekhin \ea 1998, 1999b, and references therein).
For other species, this work still remains to be done.

In the past, a lot of work has also been devoted to evaluating the conditions 
under which magnetic molecular chains may become stable in the surface layers;
furthermore, in superstrong fields they may form a magnetically stabilized
lattice 
(e.g., Ruderman 1971; Lai \& Salpeter 1997). 
Such a phase transition is also expected to have observable consequences, 
but the field still remains largely unexplored. 

\subsection{Electron gas in magnetic field}
Motion of free electrons perpendicular to the magnetic field
 is quantized  in Landau orbitals
with a characteristic transverse scale equal to
the {\it magnetic length\/} $\am=(\hbar c/eB)^{1/2}=\Compton/\sqrt{b}$,
where $b=\hbar\wc/m_ec^2 = B_{12}/44.14$ 
is the magnetic field strength  expressed in relativistic units,
$\wc={eB/mc}$ is the electron cyclotron frequency,
and $B_{12}\equiv B/10^{12}$~G.
The Landau energy levels are
\begin{equation}
   \epsilon = \epsilon_n(p_z) = c\,\left(m_e^2 c^2 + 2\hbar\wc
   m_e n+p_z^2\right)^{1/2},
\label{magnenergy}
\end{equation}
with $n=0,1,2,....$, where the magnetic field 
$\vB$ is assumed to be homogeneous and 
directed along the $z$ axis, and $p_z$ is the longitudinal momentum.
The ground Landau level $n=0$ is non-degenerate with respect
to spin projection ($s=-1$, statistical weight $g_0=1$)
while the levels $n>0$ are doubly degenerate ($s=\pm1$, $g_n=2$).
The anomalous magnetic moment of the 
electron, $g_{\rm e} = 1.00116$, 
causes splitting of the energy levels $n\geq1$ by
$\delta \epsilon = (g_{\rm e}-1) \hbar\wc$,
which, strictly speaking, removes the double spin-degeneracy.
In typical NS envelopes, this splitting is 
negligible because $\delta\epsilon$ is
smaller than either the thermal width $\sim k T$
or the collisional width of the Landau levels.

The electron's phase space is thus now a combination of an 
energy continuum in $p_z$, corresponding to the motion along the field, 
and a discrete spectrum (the quantum number $n$) 
corresponding to the quantized  transverse motion. 
This property will be reflected in most of the physical processes 
of our interest here. 

Let us denote by $n^{*}$ the highest Landau excitation populated at a 
given energy $\epsilon$. It equals an integer part of the combination
$p_0^2(\epsilon)/(2\mel\hbar\wc)$, 
where
$
   p_n(\epsilon) = [(\epsilon/c)^2 - (\mel c)^2
           - 2 \mel \hbar \wc n]^{1/2} = |p_z|.
$
Taking into account that the number of quantum states of an electron
with given $s$ and $n$ 
in volume $V$ per longitudinal momentum interval $\Delta p_z$
equals $V\Delta p_z/(4\pi^2 a_m^2 \hbar)$,
one can obtain the electron number 
density and pressure from first principles
(e.g., Blandford \& Hernquist 1982).
For strongly degenerate electrons,
\begin{eqnarray}
  n_{\rm e}&=& {1\over 2 \pi^2 a_m^2 \hbar}
            \sum_{n=0}^{n^*} g_n p_n(\EF),
\label{n-mag}
\\
   P_{\rm e} &=& P_0\,{b\over 4\pi^2}\,
       \sum_{n=0}^{n^*} g_n\,(1+2b n)
        \,[x_n\sqrt{1+x_n^2}-\ln(x_n+\sqrt{1+x_n^2})],
\label{P-mag}
\end{eqnarray}
where $x_n=c p_n(\EF)/\epsilon_n(0)$,
and $P_0$ is the same as in Eq.~\req{P-d}.

For a degenerate electron gas the  
thermodynamic quantities such as pressure, magnetization, and energy density
exhibit quantum oscillations of the de Haas--van Alphen type 
whenever the dimensionless Fermi momentum reaches the characteristic values 
$\chi=\sqrt{2nb}$ which signify the occupation of new Landau levels. 
In these oscillations the various quantities typically take values 
around their classical $B=0$ values, except in the
limit of a {\it strongly quantizing\/} field ($n^{*}=0$)
where one often finds substantial deviations
(e.g., Yakovlev \& Kaminker 1994).
The latter case takes place 
when the typical energies $kT,~kT_{\rm F} < \epsilon_1-1$
-- i.e.,
at $T\ll T_B$
and $\rho<\rho_B$,
where
\begin{eqnarray}
    \rho_B =  m_{\rm u} n_B \mu_{\rm e}
 & \approx& 7045
       \,B_{12}^{3/2}\,\mu_{\rm e}\gcc,
\label{rho_B}
\\
    T_B = 
    {\hbar\omega_c/k\GF} &\approx &
             {1.343\times10^8\, (B_{12}/\GF)}{\rm~K},
\label{T_B}
\end{eqnarray}
and $n_B=1/(\pi^2\sqrt2\,\am^3)$ is the electron number density
at which the Fermi energy reaches the first excited Landau level.
This case is of special  
interest for the NS outer layers under consideration. 

\paragraph{Strongly quantizing field.}
When the electron's transverse 
motion is frozen in the ground state Landau level 
$n^{*}=0$, the phase-space is effectively one-dimensional. 
Then $\epsilon=c\,\sqrt{(m_e c)^2 + p_z^2}$,
and Eq.~\req{n-mag} simplifies to
\beq
\quad p_{\rm F}/\hbar=2\pi^2 a_m^2 n_{\rm e}. 
\label{pF-mag}
\eeq 
We therefore see that the dimensionless Fermi momentum, 
\begin{equation}
 \chi\equiv{p_{\rm F}/{m_ec}}=(2/3)\,\chi_0^3/b
     \approx(0.6846/b) \,\rho_6/\mu_{\rm e},     
 \label{chi-mag} 
\end{equation}
is proportional to the density $\rho$, 
in sharp contrast to the non-magnetic Eq.~\req{x_r}.
Henceforth we denote $p_{\rm F0}=\mel c\chi_0$
the ``classical'' (non-magnetic)
Fermi momentum at a given density,
and reserve notation $p_{\rm F}=\mel c\chi$
for the same quantity modified by the magnetic field.
According to Eq.~\req{rho_B}, the strongly quantizing regime
in which Eq.~\req{chi-mag} is valid requires $\chi<\sqrt{2b}$.

The Fermi temperature $T_{\rm F}$ is again given by Eq.~\req{T_F}, 
but with the modified $\chi$.
 Since
$
    \chi = (4/3)^{1/3} ({\rho/\rho_B})^{2/3}
      \chi_0,
$
$T_{\rm F}$ is strongly reduced
at $\rho\ll\rho_B$. Conversely, at a given $T<T_B$,
the degeneracy takes hold at much higher density than in the $B=0$ case. 
An initially degenerate electron gas at $B=0$ will thus become non-degenerate 
when a strong quantizing field is switched on.

 Let us now consider the EOS.
Since $n=0$, Eq.~\req{P-mag} simplifies considerably.
Given Eq. \req{chi-mag}, this expression again takes the form of a power-law 
of the density, 
$P=P_0 b \chi^{\gamma_{\rm ad}}/(2\pi^2{\gamma_{\rm ad}})
      \propto\rho^{\gamma_{\rm ad}}/B^{\gamma_{\rm ad}-1}$, 
with ${\gamma_{\rm ad}}=3$ or 2
 in the non-relativistic and ultrarelativistic limits,
respectively.

\subsection{Magnetic opacities}
In the presence of a magnetic field, the conductivity $\kappa$
becomes a tensor, so that the heat fluxes
along and across the field become different.
Since the field varies over the NS surface,
the heat transport becomes two-dimensional.
Fortunately, since the crust thickness is relatively small,
the one-dimensional equation \req{Flx} remains a good 
approximation, with $\kappa=
\kappa_\|\cos^2\theta+\kappa_\perp\sin^2\theta$,
where $\kappa_\|$ and $\kappa_\perp$ are 
the conductivities along and across $\vB$
and $\theta$ is the angle between $\vB$ and the normal to the surface.

\paragraph{Radiative opacities.}
In a magnetized plasma,
two propagating polarization normal modes are defined in the presence 
of an external field having widely different mean free paths each in the 
various photon-electron interactions (e.g., M\'esz\'aros 1992; Pavlov \ea 1995). 
Silant'ev \& Yakovlev (1980) have calculated the Rosseland opacities
for the cases when they are determined by the
Thomson and free-free processes.
When $T_B\gg T$, $\kappa_{\rm r}$ grows proportionally to $(T_B/T)^2$
-- i.e., $K_{\rm r}$ decreases as $(T/T_B)^2$.
In particular, at $T\ll T_B$, the free-free opacities tabulated
by Silant'ev \& Yakovlev (1980) tend to 
\beq
     K_{\rm r}(B)\approx ({23.2\,T / T_B})^2 K_{\rm r}(0)
       \simeq 2.2\,\bar{g}_{\rm eff}\,(Z/\mu_{\rm e}^2)\,\rho\,T_6^{-1.5}\,
          B_{12}^{-2}
{\rm~{cm^2/ g}},
\label{K-mag}
\eeq
where $K_{\rm r}(0)$ is given by Eq.~\req{K_0}.
This estimate may be used if only $T_6\la B_{12}$, $\chi\ll1$,
and the opacity is dominated by the free-free processes.

{}From Eqs.~\req{P-nondeg}, \req{KP}, and \req{K-mag} one sees immediately
that the {\it radiative surface\/} in the strong magnetic
fields, $B_{12}\ga T_6$,
 is pushed to the higher densities, $\rho_{\rm s}\propto B$.

Caution is necessary however while using the scaling law \req{K-mag}.
The strong magnetic field shifts the ionization equilibrium
toward a lower degree of ionization, because of the
increasing binding energies. Therefore, 
even if the plasma were fully ionized at some $\rho$ and $T$
in the absence of magnetic field, it may be only partially ionized
at the same $\rho$ and $T$ when $B$ is high.
This increases the significance of the bound-bound and bound-free opacities
and may result in a total radiative opacity 
considerably larger than that given by Eq.~\req{K-mag}
(see, e.g., Potekhin \ea 1998).

\paragraph{Electron conductivities.}
Unified expressions and fitting formulae for thermal and electrical
electron conductivities in a fully ionized degenerate plasma
with arbitrary magnetic field have been obtained recently
(Potekhin 1999).
These conductivities undergo oscillations of the de Haas--van Alphen
type at $\rho\ga\rho_B$. At $B\gg10^{10}$~G,
the transport across the field is suppressed by orders of
magnitude. This fact allows us to neglect $\kappa_\perp$ totally,
which will be a good approximation everywhere
except at a narrow stripe near the magnetic equator, 
where $\theta\approx\pi/2$.

\begin{figure}[t]
\epsfysize=75mm
\epsffile[30 220 550 590]{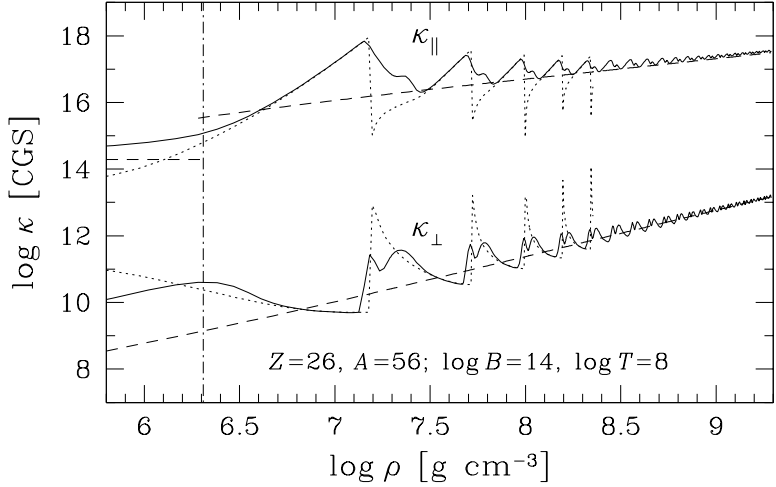}
\caption{Longitudinal ($\|$) and transverse ($\perp$)
thermal conductivities  in the outer NS envelope composed of iron at 
$T=10^8$~K and $B=10^{14}$~G: comparison of accurate results (solid lines)
with the classical approximation (dashed lines)
and with the results without thermal averaging.
The electrons are degenerate to the right of
the vertical dot-dashed line, which corresponds to $T=T_{\rm F}$.
The dashed horizontal line in the non-degenerate region
shows the conductivity given by Eq.~\protect\req{kappa-ND}
with $F_Z/\Lambda=1$.
}
\label{fig-cond}
\end{figure}

This approximation, $\kappa\approx\kappa_\|\cos^2\theta$,
 holds despite the arguments by
Heyl \& Hernquist (1998) that $\kappa_\perp$ 
becomes non-negligible in the strongly quantizing regime
at low density because the ratio 
$\kappa_\perp/\kappa_\|$ evaluated for strongly degenerate electrons
increases without bound as $\chi\to0$. However, 
the finite thermal width of the Fermi level (neglected
by these authors) removes the divergency.
At typical NS temperatures, the thermal averaging
terminates the growth of $\kappa_\perp$ and moderates the decrease
of $\kappa_\|$, before they become comparable.
In Fig.~\ref{fig-cond}, we plot
by solid lines an example of the thermal conductivities
calculated according to Potekhin (1999).
For comparison, the dotted curves display
the conductivities which would have been obtained without
thermal averaging. 
As the electrons become non-degenerate at low densities, the 
solid curve for $\kappa_\|$ is seen to level off. 
It tends to its non-degenerate value 
$\kappa_\|^{\rm nd}$ which 
is of the order of the non-magnetic value, Eq.~\req{kappa-ND},
depending on density only logarithmically.

The dashed lines show the ``classical'' approximation where
the quantizing nature of the field is neglected.
For $\kappa_\|$, this approximation is close to the non-magnetic one.
We can see that at high enough densities
beyond the first 
oscillation, the classical approximation
is good enough. At lower densities, the quantizing nature of the field
must be taken into account. At $\rho<\rho_B$,
the neglect of thermal averaging is justified
as long as the electrons are degenerate.
Then the longitudinal conductivity may be written as
\beq
   \kappa_\| = {k^2 T\, (\mel^2 c^3 b)^2
          \over 12\pi\hbar^3 Z e^4 n_{\rm e}}\,{\chi^2\over2Q^\|}
       = \frac23\,{\kappa_0\,(1+\chi_0^2)\,\Lambda\over Q^\|},
\eeq
where $\kappa_0$ is the non-magnetic conductivity
given by Eq.~\req{K_ei},
in which $\chi$ must be replaced by $\chi_0$,
$\Lambda$ is the non-magnetic Coulomb logarithm, 
and $Q^\|$ is a function of $\chi$ defined by Eq.~(A9)
in Potekhin (1999).
In the liquid regime (far enough from the solid phase boundary),
the latter function reduces to the expression (Yakovlev 1984)
\beq
   Q^\| = \xi^{-1}-{\rm e}^\xi E_1(\xi),
\label{Q}
\eeq
where $\xi=2\chi^2/b+\frac12(\am q_{\rm s})^2$,
$q_{\rm s}$ is an effective Coulomb-screening wave number,
and $E_1$ is the standard exponential integral.
A simple order-of-magnitude estimate of $\kappa_\|$
in the degenerate Coulomb liquid can be obtained 
if we neglect $q_{\rm s}$ and the second term in Eq.~\req{Q}. In this way
we obtain
\beq
   \kappa_\|\simeq(4/3)\,\kappa_0\,\Lambda(1+\chi_0^2)\,\chi^2/b
      \simeq 5\times10^{15}\,Z^{-1}\,T_6\,\chi^3
 {\rm~erg/(cm~s~K)}.
\label{kappa-mag}
\eeq

As noted earlier, in the very strong fields considered here the onset of 
degeneracy is pushed into ever increasing densities as $B$ increases.
Therefore the turnover from radiative heat transfer to 
electron-conduction dominated transport may occur 
in the non-degenerate regime. 
In this case, Eq.~\req{kappa-ND} may be used 
to evaluate $\kappa_\|$. 

\subsection{Consequences for the heat transport}
We have seen that in strong magnetic fields
there are several different regimes 
regulating the EOS and opacities.
For the construction of an approximate analytic
thermal profile it is sufficient to note that
the non-magnetic expressions for the radiative opacity and 
{\it longitudinal\/} electron
conductivity $\kappa_\|$ remain good approximations
 unless the field is strongly quantizing.
Magnetic oscillations, which occur around the classical functions,
will be smoothed out by integration while obtaining
the thermal profile from Eq.~\req{Flx};
thus they are not too important.

When the field is strongly quantizing,
the opacities are modified appreciably.
However, in the  degenerate part of 
NS ocean, which is of our prime interest here,
the analytic expressions for $\kappa_\|$
can be again approximated as a power law, Eqs.~\req{kappa-mag} and 
\req{kappa-ND}.
The same is true with respect to 
the extreme quantizing limit of radiative opacity,
as follows from Eq.~\req{K-mag}.
In the non-degenerate regime, the magnetic field does not affect the EOS.
In this case, we recover the solution \req{prlw}
with new values of $\beta=-1.5$ and $K_0$, given by Eq.~\req{K-mag}. 
Then
\beq
   T_6\approx 0.95\, (\bar{g}_{\rm eff}\,q)^{2/9}
            \,(\rho/\mu_{\rm e})^{4/9}\,B_{12}^{-4/9},
\label{nd-mag}
\eeq
with the same $q$ as in Eq.~\req{nd-solution}.
Thus the temperature is reduced (its profile becomes less steep)
with increasing $B$ as long
as the field is strongly quantizing ($\rho<\rho_B$).

It is interesting to note that the value of the constant conductivity, 
Eq.~\req{k-const} is independent of the magnetic field, while its
numerical 
value is only slightly lowered as a result of the changed coefficient 
$\beta$.

\paragraph{Sensitivity strip.} As we have seen, the sensitivity strip
is placed near the turning point from radiative transport 
to electron conduction, defined by $K_{\rm r}=K_{\rm c}$.
In a strongly quantizing field, using Eqs.~\req{K-mag} and
\req{kappa-mag},
we have 
\beq
\rho\approx 250\,\mu_{\rm e}\,\bar{g}_{\rm eff}^{-0.2}\, T_6^{0.7}\,
|\cos\theta|^{-0.4} B_{12}~\gcc
\label{turn-mag}
\eeq
instead of Eq.~\req{turn1}. 
With the temperature profile \req{nd-mag},
we now obtain
\beq
     T_{\rm t}\approx {3.5\times10^7\over|\cos\theta|^{8/31}}
     \,\bar{g}_{\rm eff}^{6/31}\,
        q^{10/31}\,{\rm~K},
\quad
   {\rho_{\rm t}\over \mu_{\rm e}}
     \approx 
      {3\times10^3 \, q^{7/31}\,B_{12}
        \over
          |\cos\theta|^{18/31}\,\bar{g}_{\rm eff}^{2/31}}\gcc.
\eeq

If, however, the electrons are non-degenerate 
along the turnover line, then $\kappa_\|$ 
is represented by Eq.~\req{kappa-ND} instead of \req{kappa-mag}.
In this case, we obtain the 
turnover at 
\beq
\rho\approx 52\,\sqrt{\Lambda/F_Z}\,\mu_{e}\,\bar{g}_{\rm eff}^{-1/2}\, T_6\,
|\cos\theta|^{-1}\,B_{12}~\gcc.
\label{turn-nd}
\eeq
Applying Eq.~\req{nd-mag}, we get
\[
T_{\rm t}\approx {2.2\times10^7\over|\cos\theta|^{0.8}}
\left({\Lambda q\over F_Z}\right)^{0.4}{\rm K},
\quad
   {\rho_{\rm t}\over \mu_{\rm e}}
     \approx 
      {1.1\times10^3 \, q^{0.4}\,B_{12}\over
          |\cos\theta|^{1.8}\,\bar{g}_{\rm eff}^{0.5}}
\left({\Lambda\over F_Z}\right)^{0.9}
\gcc.
\]

Thus, in both cases (for degenerate and non-degenerate electrons)
we have obtained similar dependences of the position of the
turning point on the NS parameters.
These dependences should be compared with Eq.~\req{turn}.
We see that $T_{\rm t}$ in both equations have similar values
at $\theta=0$, but in the magnetic field
$T_{\rm t}$ increases with increasing $\theta$. 
It is noteworthy that $T_{\rm t}$ is independent of $B$, 
while $\rho_{\rm t}$
grows linearly with $B$ in the strongly quantizing field.
{}From Eq.~\req{rho_B} we can evaluate the condition
at which $\rho_{\rm t}$ lies in the
region of strong magnetic quantization.
Assuming that $\theta$ is not close to $\pi/2$ and
neglecting factors about unity, we see that $\rho_{\rm t}<\rho_B$ for 
$B_{12}\ga (Z\,T_{\rm e6}^4/g_{14})^{14/31}$,
i.e., for the strong-field pulsars and magnetars.

\paragraph{Onset of degeneracy.} 
Let us estimate the point at which the
electrons become degenerate. For simplicity,
let us assume that the electrons are non-relativistic.
Taking into account Eqs.~\req{T_F} and \req{chi-mag}, we see
that the condition $T=T_{\rm F}$ in the strongly quantizing magnetic field
is equivalent to $\rho\approx608\,\mu_{\rm e}\,\sqrt{T_6}\,B_{12}$.
Then from Eq.~\req{nd-mag} we obtain
$
\rho_{\rm d}\simeq3700\,(\bar{g}_{\rm eff}\,q)^{1/7} \mu_{\rm e} B_{12}.
$
Thus, similar to the non-magnetic case, the switch
between the regimes of photon and electron heat conduction
occurs not far from the onset of degeneracy: $\rho_{\rm d}\sim\rho_{\rm t}$. 
Depending on $\theta$, it can occur either in 
the non-degenerate domain (at $\theta\approx0$)
or in the degenerate domain (at $\theta\ga60^\circ$).

 \begin{figure}[t]
 \epsfysize=57mm
 \epsffile[60 450 570 680]{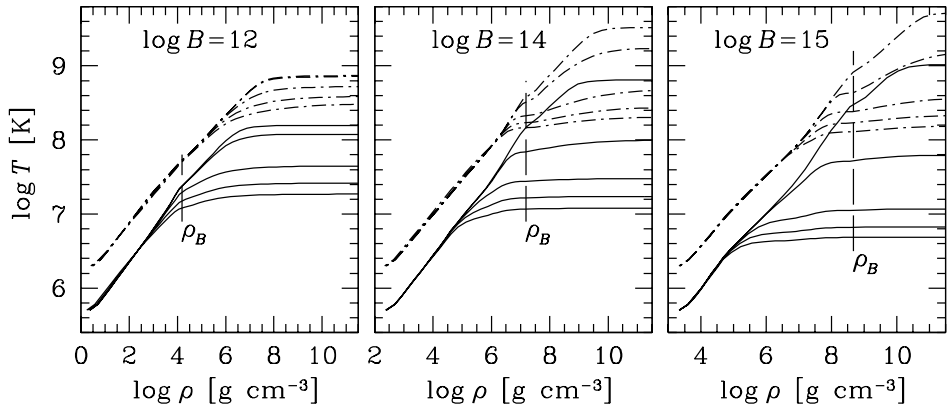}
 \caption{Temperature profiles through an iron envelope of
 a NS with $M=1.4\,M_\odot$, $R=10$ km,
 and with magnetic field $B=10^{12}$ (left), $10^{14}$ (middle),
 or $10^{15}$~G (right panel).
 In every case,
 effective surface temperature was fixed to $T_{\rm e}=5\times10^5$~K
 (solid lines) or $2\times10^6$~K (dot-dashed lines).
 Different lines of each bunch correspond to 
 different values of $\cos\theta$: 1 (the lowest line of a bunch), 
 0.7, 0.4, 0.1, and 0 (the highest line).
 }
 \label{fig-profmag}
 \end{figure}

\paragraph{Beyond the turning point.}
The integration beyond the turning point in order to obtain $T_{\rm i}$
can be done in the same way as in the non-magnetic case.
However, since the turnover occurs in the strongly quantizing field,
the integration path should be divided in two parts:
below $\rho_B$, where Eq.~\req{kappa-mag}
should be used for the conductivity, and above $\rho_B$,
where one can use Eq.~\req{dT/dx} with the right-hand side
divided by $\cos^2\theta$. The result is similar to
Eq.~\req{T1}, but contains a profound dependence on 
the inclination angle: the thermal gradient
grows rapidly as $\theta$ increases toward $\pi/2$.

 This dependence is illustrated in Fig.~\ref{fig-profmag},
 where we have shown several temperature profiles
 calculated numerically (Potekhin 2000).
 The curves start at the radiative surface, where $T=T_{\rm e}$,
 and end near the neutron drip point.
 The onset of the profound $\theta$-dependence signals
 the turning point. According to our estimates above,
this point 
 is shifted to ever higher densities with increasing $B$.
For the radiative boundary, the above-mentioned
linear dependence $\rho_{\rm s}\propto B$
is confirmed.
 Another interesting effect is that $T$ may continue to grow
 up to the bottom of the outer envelope,
well beyond the ``canonical'' limiting density $\rho=10^{10}\gcc$.
 
\section{Conclusions}
In the preceding sections we have given a simplified analytic discussion 
of the neutron star outer envelope. It was possible in this simple picture 
to recover main features of more careful numerical calculations,
such as the sensitivity of the temperature ratio 
$T_{\rm i}/T_{\rm e}$ to chemical 
constitution and the general behavior of the 
$(T, \rho)$ profile. 

Magnetic fields have a strong influence on the radiative and conductive 
opacities in NS envelopes, as we have seen in the last section. 
It is remarkable though that
in the region of the magnetic polar cap, in spite of the sharply reduced 
radiative opacities in the outer layers, the temperature ratios 
$T_{\rm t}/T_{\rm e}$, and $T_{\rm i}/T_{\rm e}$ are only mildly affected. 
The effect of the reduced opacity is partly counterbalanced by the 
increased mass of the radiative layer. 
Furthermore, when the magnetic field is inclined 
to the surface, the temperature gradient is increased
because the electron conduction
is efficient mainly along the field lines.
Clearly more work is still necessary to fully analyze all these details. 

The possibility of magnetars, or extremely highly magnetized NS, in our 
Galaxy has been highlighted by recent observations of soft gamma repeaters
(Kouveliotou 1998, 1999; Thompson 2000) and the special class of 
anomalous X-ray pulsars (e.g., Mereghetti 2000). 
Understanding the role of high magnetic fields in the observable
properties 
of such objects has thus become an important theoretical task. 
Recent observations of compact galactic objects further suggest 
rather small emission areas and high temperatures, which may be 
attributed to magnetars (e.g., Pavlov \ea 2000; Dar \& de R\'ujula 2000). 
The presence of an accreted hydrogen layer on the magnetar's polar cap 
could allow for a rather sharp temperature contrast between the hot spot
and 
the rest of the NS surface. This configuration has been invoked 
to interpret the observations in the case of the central compact object in
the supernova remnant Cas A (Pavlov \ea 2000). 

\paragraph{Acknowledgments.}
We thank D.G.\ Yakovlev for useful remarks.
J.V. is deeply indebted to Vladimir Usov and Motty Milgrom for their
generous hospitality and partial support from the Einstein Center for Theoretical
Physics at the Weizmann Institute of Science; he also acknowledges partial 
support from EC grant ERB-FMRX-CT98-0195.
A.Y.P.\ gratefully acknowledges generous hospitality of Gilles Chabrier
at CRAL, Ecole Normale Sup\'erieure de Lyon,
and partial support from grants
RFBR 99-02-18099 and INTAS 96-542.



\begin{thebibliography}{}  
\small
\bibitem[]{}
Alpar, M.A., Cheng, K.S., \& Pines, D. (1989).
Vortex creep and the internal temperature of neutron stars,
\apj346,823         
\bibitem[]{}
Braginski\u\i, S.I. (1957).
Transport phenomena in a plasma,
{\it Zh.\ Eksp.\ Teor.\ Fiz.} {\bf 33}, 645
\bibitem[]{}
Blandford, R.D., \& Hernquist, L. (1982).
Magnetic susceptibility of a neutron star crust,
{\it J.\ Phys.\ C: Solid State Phys.} {\bf 15}, 6233
\bibitem[]{}
Canuto, V., \& Ventura, J. (1977).
Quantizing magnetic fields in astrophysics,
{\it Fundam.\ Cosm.\ Phys.} {\bf 2}, 203
\bibitem[]{}
Caraveo, P., Bignami, G. \& Tr\"umper, J. (1996).
Radio-silent isolated neutron stars as a new astronomical reality,
\aar7,209
\bibitem[]{}
Chabrier, G. (1993).
Quantum effects in dense Coulombic matter --
Application to the cooling of white dwarfs,
\apj414,695
\bibitem[]{}
Chabrier, G., \& Potekhin, A.Y. (1998).
Equation of state of fully ionized electron-ion plasmas,
\pre58,4941
\bibitem[]{}
Chiu, H.Y., \& Salpeter, E.E. (1964).
Surface X-ray emission from neutron stars,
\prl12,413
\bibitem[]{}
Cox, J.P., \& Giuli, R.T. (1968).
{\it Principles of Stellar Structure}
 (Gordon and Breach, New York)
\bibitem[]{}
Dar, A. \& de R\'ujula, A. (2000).
SGRs and AXPs: Magnetars or young quark stars?
astro-ph/0002014
\bibitem[]{}
Gudmundsson, E.H., Pethick, C.J., \& Epstein, R.I. (1983).
Structure of neutron star envelopes, 
\apj272,286 (GPE)
\bibitem[]{}
Halpern, J.P., \& Ruderman, M. (1993).
Soft X-ray properties of the Geminga pulsar, 
\apj415,286
\bibitem[]{}
Hernquist, L., \& Applegate, J.H. (1984).
Analytical models of neutron star envelopes, 
\apj287,244 (HA84)
\bibitem[]{}
Heyl, J.S., \& Hernquist, L. (1998).
Almost analytic models of ultramagnetized neutron star envelopes, 
 \mn300,599
\bibitem[]{}
Iglesias, C.A., \& Rogers, F.J. (1996).
Updated OPAL opacities,
\apj464,943; \verb|http://www-phys.llnl.gov/V_Div/OPAL/opal.html|
\bibitem[]{}
Kouveliotou, C. {\it et al.\/} (1998).
An X-ray pulsar with a superstrong magnetic field in 
the soft gamma-ray repeater SGR 1806--20,
\nat393,235
\bibitem[]{}
Kouveliotou, C. {\it et al.\/} (1999).
Discovery of a magnetar associated with 
the soft gamma repeater SGR 1900+14, 
\apj510,L115
\bibitem[]{}
Lai, D., \& Salpeter, E.E. (1997).
Hydrogen phases on the surface of a strongly magnetized 
neutron star,
\apj491,270
\bibitem[]{}
Mereghetti S. (2000).
The anomalous X-ray pulsars,
this volume
\bibitem[]{}
M\'esz\'aros, P. (1992). 
{\it High Energy Radiation from Magnetized Neutron Stars}
(U. of Chicago Press, Chicago)
\bibitem[]{}
Miller, M.C., \& Neuhauser, D. (1991).
Atoms in very strong magnetic fields,
\mn253,107
\bibitem[]{}
Page, D. (1997).
Fast cooling of neutron stars, 
\apj479,L43
\bibitem[]{}
Page, D., \& Sarmiento, A. (1996).
Surface temperature of a magnetized neutron star and interpretation 
of the \ros  data. II, 
 \apj473,1067
\bibitem[]{}
Pavlov, G.G., Shibanov, Yu.A., Zavlin, V.E., \& Meyer, R.D. (1995). 
Neutron star atmospheres, in: M.A. Alpar, \"U. Kizilo\u{g}lu, 
\& J. van Paradijs (eds.),
{\it The Lives of the Neutron Stars},
NATO ASI Ser. C {\bf 450} 
(Kluwer, Dordrecht) p.~71
\bibitem[]{}
Pavlov, G.G., Zavlin, V.E., Aschenbach, B., Tr\"umper, J., \& Sanwal, D.
(2000). 
The compact central object of Cassiopeia A:
a neutron star with hot polar caps or a black hole?
\apj531,L53
\bibitem[]{}
Pethick, C.J. (1992).
Cooling of neutron stars,
{\it Rev.\ Mod.\ Phys.} {\bf 84}, 1133
\bibitem[]{}
Potekhin, A.Y. (1999).
Electron conduction in magnetized neutron star envelopes, 
 \aa351,787 
\bibitem[]{}
Potekhin, A.Y. (2000).
Heat and charge transport in envelopes 
of weakly and strongly magnetized neutron stars,
in: M.~Kramer, N.~Wex, \& R.~Wielebinski (eds.),
   {\it   Pulsar Astronomy -- 2000 and beyond},
      ASP Conf. Ser. {\bf 202}, 621
\bibitem[]{}
Potekhin, A.Y., Chabrier, G., \& Yakovlev, D.G. (1997).
Internal temperatures and cooling of neutron stars 
with accreted envelopes, 
 \aa323,415 (PCY)
\bibitem[]{}
Potekhin, A.Y., Shibanov, Yu.A., \& Ventura, J. (1998).
Equation of state and opacities for hydrogen atmospheres 
of strongly magnetized cooling neutron stars,
in: N.~Shibazaki, N.~Kawai, S.~Shibata, \& T.~Kifune (eds.),
{\it Neutron Stars and Pulsars}
(Universal Academy Press, Tokyo) p.~161
\bibitem[]{}
Potekhin, A.Y, Baiko, D.A., Haensel, P., \& Yakovlev, D.G. (1999a).
Transport properties of degenerate electrons 
in neutron star envelopes and white dwarf cores,
\aa346,345
\bibitem[]{}
Potekhin, A.Y., Chabrier, G., \& Shibanov, Yu.A. (1999b).
Partially ionized hydrogen plasma
in strong magnetic fields,
\pre60,2193
\bibitem[]{}
Rajagopal, M., Romani, R., \& Miller, M.C. (1997).
Magnetized iron atmospheres for neutron stars,
\apj479,347
\bibitem[]{}
Reisenegger, A. (1997).
Constraining dense matter superfluidity through thermal emission from millisecond pulsars,
\apj485,313
\bibitem[]{}
Romani, R.W. (1987).
Model atmospheres for cooling neutron stars,
\apj313,718
\bibitem[]{}
Ruderman, M.A. (1971).
Matter in superstrong magnetic fields:
the surface of a neutron star,
\prl27,1306
\bibitem[]{}
Shapiro, S.L., \& Teukolski, S.A. (1983).
{\it Black Holes, White Dwarfs, and Neutron Stars: The Physics of Compact 
 Objects} (Wiley, New York) (ST83)
\bibitem[]{}
Shibanov, Yu.A., Potekhin, A.Y., Yakovlev, D.G., \& Zavlin, V.E. (1998).
Evolution of thermal structure and radiation spectrum of cooling neutron stars,
in: R. Buccheri, J. van Paradijs, \& M.A. Alpar (eds.),
     {\it The Many Faces of Neutron Stars}
     (Kluwer, Dordrecht) p. 553 
\bibitem[]{}
Silant'ev, N.A., \& Yakovlev, D.G. (1980).
Radiative heat transfer in surface layers of neutron stars 
with a magnetic field,
\ass71,45
\bibitem[]{}
Thompson, C. (2000).
The soft gamma repeaters,
this volume
\bibitem[]{}
Thorne, K.S. (1977).
The relativistic equations of stellar structure and evolution,
\apj212,825
\bibitem[]{}
Tr\"umper, J., \& Becker, W. (1997).
X-ray emission from isolated neutron stars,
{\it Adv.\ Space Res.} {\bf 21}, 203
\bibitem[]{}
Tsuruta, S. (1998).
Thermal properties and detectability of neutron stars. II.
Thermal evolution of rotation-powered neutron stars,
{\it Phys.\ Rep.} {\bf 292}, 1
\bibitem[]{}
Urpin, V.A., \& Yakovlev, D.G. (1980).
On temperature growth inside neutron stars,
{\it Astrophysics} {\bf 15}, 429
\bibitem[]{}
Ventura, J. (1989).
Radiation from cooling neutron stars, in: H.~\"Ogelman \& E.P.J. van 
den Heuvel (eds.), 
 {\it Timing Neutron Stars} (Kluwer, Dordrecht) p. 491
\bibitem[]{}
Yakovlev, D.G. (1984).
Transport properties of the degenerate electron gas 
of neutron stars along the quantizing magnetic field,
\ass98,37
\bibitem[]{}
Yakovlev, D.G., \& Kaminker, A.D. (1994).
Neutron star crusts with magnetic fields, in: G.~Chabrier \& E.~Schatzman
 (eds.), {\it The Equation of State in Astrophysics}, Proc. IAU Coll. 147 
 (Cambridge U. Press, Cambridge, UK) p.~214
\bibitem[]{}
Yakovlev, D.G., \& Urpin, V.A. (1980).
Thermal and electrical conductivity in white dwarfs and neutron stars,
{\it Sov.\ Astron.} {\bf 24}, 303
\bibitem[]{}
Yakovlev, D.G., Levenfish, K.P., \& Shibanov, Yu.A. (1999).
Cooling neutron stars and superfluidity in their interior,
{\it Physics--Uspekhi} {\bf 42}, 737
\normalsize
\end{thebibliography}
\end{document}